\UseRawInputEncoding
\documentclass[twocolumn,superscriptaddress,floatfix]{revtex4}
\usepackage{graphicx,amsfonts,amssymb,amsmath,hyperref,enumerate,footnote}
\usepackage{graphicx}
\usepackage{float}
\usepackage{indentfirst}
\usepackage{verbatim}
\usepackage{xcolor}
\usepackage{tikz}
\usetikzlibrary{quantikz}
\usetikzlibrary{positioning}
\usepackage{array}
\usepackage[utf8]{inputenc}
\usepackage{mathtools}
\usepackage{blindtext}
\usepackage{pdflscape}
\usepackage{xcolor}
\usepackage{tikz}
\usepackage{algorithm}
\usepackage{algpseudocode}
\usetikzlibrary{arrows.meta,
                chains,
                positioning,
                shapes.geometric}
\usepackage{siunitx}
\tikzstyle{startstop} = [rectangle, rounded corners, minimum width=3cm, minimum height=1cm,text centered, draw=black, fill=red!30]
\tikzstyle{io} = [trapezium, trapezium left angle=70, trapezium right angle=110, minimum width=3cm, minimum height=1cm, text centered, draw=black, fill=blue!30]
\tikzstyle{process} = [rectangle, minimum width=3cm, minimum height=1cm, text centered, draw=black, fill=orange!30]
\tikzstyle{decision} = [diamond, minimum width=3cm, minimum height=1cm, text centered, draw=black, fill=green!30]
\tikzstyle{arrow} = [thick,->,>=stealth]
\usepackage{makecell}

\newif\ifhyper
\hypertrue
\ifhyper
\hypersetup{
   citecolor = {green},
   colorlinks = {true}, 
   urlcolor = {blue} 
}
\fi

\newcommand{\beq}{\begin{equation}}
\newcommand{\eeq}{\end{equation}}
\newcommand{\beqa}{\begin{eqnarray}}
\newcommand{\eeqa}{\end{eqnarray}}

\def\ket#1{\vert#1\rangle}

\def\Longarrow{\protect\@lra}
\def\@lra{\relbar\joinrel\relbar\joinrel\relbar\joinrel%
          \relbar\joinrel\rightarrow}

\begin{document}

\title{Hacking Cryptographic Protocols with Advanced Variational Quantum Attacks}

\author{Borja Aizpurua}
\affiliation{Multiverse Computing, Paseo de Miram\'on 170, E-20014 San Sebasti\'an, Spain}
\affiliation{Department of Basic Sciences, Tecnun - University of Navarra, E-20018 San Sebasti\'an, Spain}

\author{Pablo Bermejo}
\affiliation{Donostia International Physics Center, Paseo Manuel de Lardizabal 4, E-20018 San Sebasti\'an, Spain}

\author{Josu Etxezarreta Martinez}
\affiliation{Department of Basic Sciences, Tecnun - University of Navarra, E-20018 San Sebasti\'an, Spain}

\author{Rom\'an Or\'us}
\affiliation{Multiverse Computing, Paseo de Miram\'on 170, E-20014 San Sebasti\'an, Spain}
\affiliation{Donostia International Physics Center, Paseo Manuel de Lardizabal 4, E-20018 San Sebasti\'an, Spain}
\affiliation{Ikerbasque Foundation for Science, Maria Diaz de Haro 3, E-48013 Bilbao, Spain}

\begin{abstract} 

 Here we introduce an improved approach to Variational Quantum Attack Algorithms (VQAA) on crytographic protocols. Our methods provide robust quantum attacks to well-known cryptographic algorithms, more efficiently and with remarkably fewer qubits than previous approaches. We implement simulations of our attacks for symmetric-key protocols such as S-DES, S-AES and Blowfish. For instance, we show how our attack allows a classical simulation of a small 8-qubit quantum computer to find the secret key of one 32-bit Blowfish instance {with 24 times fewer number of iterations} than a brute-force attack. Our work also shows improvements in attack success rates for lightweight ciphers such as S-DES and S-AES. Further applications beyond symmetric-key cryptography are also discussed, including asymmetric-key protocols and hash functions. In addition, we also comment on  potential future improvements of our methods. Our results bring one step closer assessing the vulnerability of large-size classical cryptographic protocols with Noisy Intermediate-Scale Quantum (NISQ) devices, and set the stage for future research in quantum cybersecurity. 

\end{abstract}

\maketitle

\section{Introduction}
\label{sec1}

Today's world is all about information. Beyond the impact of other technologies, secure communications have become one of the cornerstones of modern society \cite{nguyen2015survey}. In this era more than ever, information is both an asset and a vulnerability, and therefore cryptology's relevance has escalated to unprecedented levels in history. Cryptology is the science of secure communication through encryption (secure data coding) and decryption (data decoding process). Information control is critical, and it is therefore no surprise that quantum computing has emerged as one of the key disruptive technologies in this ambit, given the potential of quantum computers to hack current cybersecurity standards \cite{scholten2024assessing}.

Currently, we reside in the Noisy Intermediate-Scale Quantum (NISQ) era, characterized by quantum computers that, while powerful, are still prone to errors (error rates are around $10^{-3}$ while $10^{-12}$ is considered to be required for fault-tolerance) and limitations in qubit coherence times \cite{preskill2018quantum,ibmrigetticoherences}. In this scenario, the community is actively working on innovations and breakthroughs to advance beyond the limitations of the NISQ era. The primary goal is to develop fault-tolerant quantum computers that can fully harness the quantum advantage. Meanwhile, significant effort is dedicated to maximizing the potential of existing NISQ devices.

The field of cryptology serves as the cornerstone of secure digital communication, encompassing both cryptography and cryptanalysis. Cryptography provides tools for encrypting sensitive data, thereby ensuring its confidentiality (protecting data privacy) and integrity (ensuring data accuracy) during transmission over insecure networks. Cryptanalysis, the counterpart to cryptography, aims to break these secure communication channels. It employs a variety of techniques, ranging from brute-force attacks (which involve an exhaustive search through all possible keys) to more structured approaches. The latter may include ciphertext-only (decryption without key knowledge), known-plaintext (decryption with known data), and chosen-plaintext attack (attacker chooses input to decrypt output), each one posing a specific set of challenges for the attacker \cite{schneier2007applied, book_crypto, hash}. These types of attacks are used by the cryptography community for establishing the security of the methods \cite{BernsteinPQC,PQCreview}.

Cryptographic functions can be divided into hash algorithms (data integrity and password verification \cite{hash}), symmetric algorithms (use a single secret key for both encryption and decryption providing speed advantages) and asymmetric algorithms (RSA \cite{RSA}, which uses a public key for encryption and a private key for decryption). For well-known symmetric ciphers such as DES \cite{des}, AES \cite{AES}, and Blowfish \cite{blowfish}, advanced classical cryptanalysis methods, such as linear cryptanalysis, biclique attacks, and related-key attacks, have been shown to outperform brute-force in terms of computational complexity \cite{matsui1993linear, biclique2011}. However, these methods often rely on exploiting specific structural weaknesses or assumptions about key biases. In contrast, our approach focuses on general-purpose quantum cryptanalysis, specifically targeting known plaintext-ciphertext attacks without leveraging protocol-specific shortcuts. This generality ensures wide applicability and complements existing classical methods, while showcasing the potential of hybrid quantum-classical strategies in cryptanalysis.

In this context, quantum computing \cite{quantum-computing} has challenged the very foundations of cryptology. By leveraging the principles of quantum mechanics, quantum  computers hold the promise of polynomial or even exponential speed-ups for certain computational tasks, including the ones often encountered in cryptography. Quantum computers therefore pose a significant threat to classical cryptographic systems. The factoring algorithm by Shor \cite{shor} showed that quantum computers could, in practice, recover the key of RSA and related cybersecurity protocols efficiently. Recent estimations suggest that around 20 million noisy qubits (error-prone quantum bits) are needed to construct an error corrected quantum algorithm (Shor) that breaks encryption of 2048-bit RSA keys in 8 hours \cite{20million}, still beyond reach of current quantum processors \cite{bharti2022noisy}. More generically, brute-force attacks can also be enhanced via quantum computing using Grover's algorithm for searching unstructured databases \cite{grover}, and which has been tested for symmetric DES protocols \cite{grover-attack}. Cryptography and cryptanalysis are critically important with the advent of quantum computers, as evidenced by the efforts of the American National Institute of Standards and Technology (NIST) to standardize Post-Quantum Cryptography (PQC) schemes. NIST's PQC initiative aims to develop cryptographic systems that are secure against both quantum and classical computers, ensuring the confidentiality and integrity of digital communications well into the future \cite{NIST-PQC}.

In the current era of NISQ devices, promising proposals have been made in the field of variational quantum algorithms (VQAs), which are designed to find the minimum of a function by iteratively adjusting the parameters of a quantum circuit encoding the solution for such a function. VQAs have found wide-ranging applications beyond cryptography, including quantum chemistry, where they have been employed for solving molecular electronic structure problems \cite{ref1}; machine learning, through approaches such as quantum neural networks and federated learning frameworks \cite{ref2, ref3}; and quantum finance, with applications to Monte Carlo pricing of financial derivatives \cite{ref4}. For example, the Variational Quantum Attack Algorithm (VQAA) \cite{VQAA} is designed to recover the key from the Simplified-Data Encryption Standard (S-DES) \cite{sdes}. This approach can be extrapolated to more complex AES-like symmetric protocols. In practice, VQAA encodes a known ciphertext in the lowest-energy state of a classical Hamiltonian (system's total energy equation model), akin to a cost function, and aims to find such state by trying out different keys, aiming to retrieve the secret key by optimizing a set of variational parameters.

The goal of this paper is two-fold. First, we show how the performance of VQAA can be dramatically boosted by (i) significantly reducing the number of qubits and circuit depth needed, and (ii) introducing coordinate transformations. We reduce the number of qubits and circuit depth by showing that qubit measurements can be performed before the encryption step, producing the same output as if they were performed after, which allows making the encryption process classical. This approach leverages non-orthogonal quantum states to encode multiple bits of information within a single qubit \cite{non-orthogonal, PerezSalinas2020datareuploading}. The generated key is obtained by calculating the reduced density matrix (RDM) of each qubit to extract the most probable state of each qubit. The number of measurements required is those needed to get the RDM of individual qubits, not the complete state, making this method scalable with the number of qubits. Reducing circuit depth and qubit count is especially relevant for VQAs in the NISQ regime, particularly to avoid Noise Induced Barren Plateaus \cite{cerezoNIBP}. We also utilize alternative cost functions, such as the Hamming distance, to improve the efficiency of the encoding. The algorithm's convergence is improved through the application of coordinate transformations to the optimization parameters \cite{rotaxis}. These transformations, similar to variational rotations, effectively mitigate issues like local minima and Barren plateaus \cite{mcclean2018barren} by adjusting the optimization landscape.

Second, we show how the same approach can be generalized to deal with essentially any type of cryptanalysis, including symmetric and asymmetric keys as well as hash functions. For the sake of concreteness, here we focus on analyzing symmetric-key protocols, showing that our methods outperform traditional brute-force analysis in breaking the security of the Blowfish cipher, as well as improve success rates in attacks to lightweight ciphers such as S-DES and Simplified-AES. It is important to note that our method is heuristic, focusing on practical improvements and empirical results rather than formal mathematical proofs. Specifically, the ansatz design and cost function selection are guided by heuristics aimed at balancing quantum expressiveness with resource constraints, ensuring scalability and applicability in cryptographic contexts. These heuristic choices are validated through numerical experimentation and tailored to address the specific challenges of cryptanalysis.

This paper is organized as follows. Sec. \ref{sec2} provides a comprehensive overview of the techniques and algorithms employed, and is further divided into multiple subsections covering different aspects of our algorithms. In Sec. \ref{sec3} we present the results of our numerical analysis for S-DES, S-AES and Blowfish. Sec. \ref{sec4} discusses generalizations of our method to asymmetric-key cryptography and hash functions. Sec. \ref{sec5} discusses further improvements to be implemented to enhance performance even further. Finally, Sec. \ref{sec6} wraps up with the conclusions and perspectives of our work. 

\section{Methodology}
\label{sec2}

Let us discuss now our method in detail. The core of our technique is inspired by an improved version of the VQAA algorithm in \cite{VQAA}. After analyzing the details of that algorithm, we will arrive to the conclusion that (i) it can be implemented with fewer qubits and circuit depth with the same exact accuracy, and (ii) it can be extended well beyond the use-case of symmetric key cryptography. In addition, we also note that the algorithm, as such, can be enhanced by a series of improvements, such as better cost functions (not necessarily quadratic Hamiltonians), non-orthogonal qubit encodings as in \cite{non-orthogonal}, and acceleration schemes  {alleviating detrimental effects} of Barren plateaus as in \cite{rotaxis}. Let us discuss this in further detail in what follows. 

\begin{figure}
\centering
  \includegraphics[width=0.8\columnwidth]{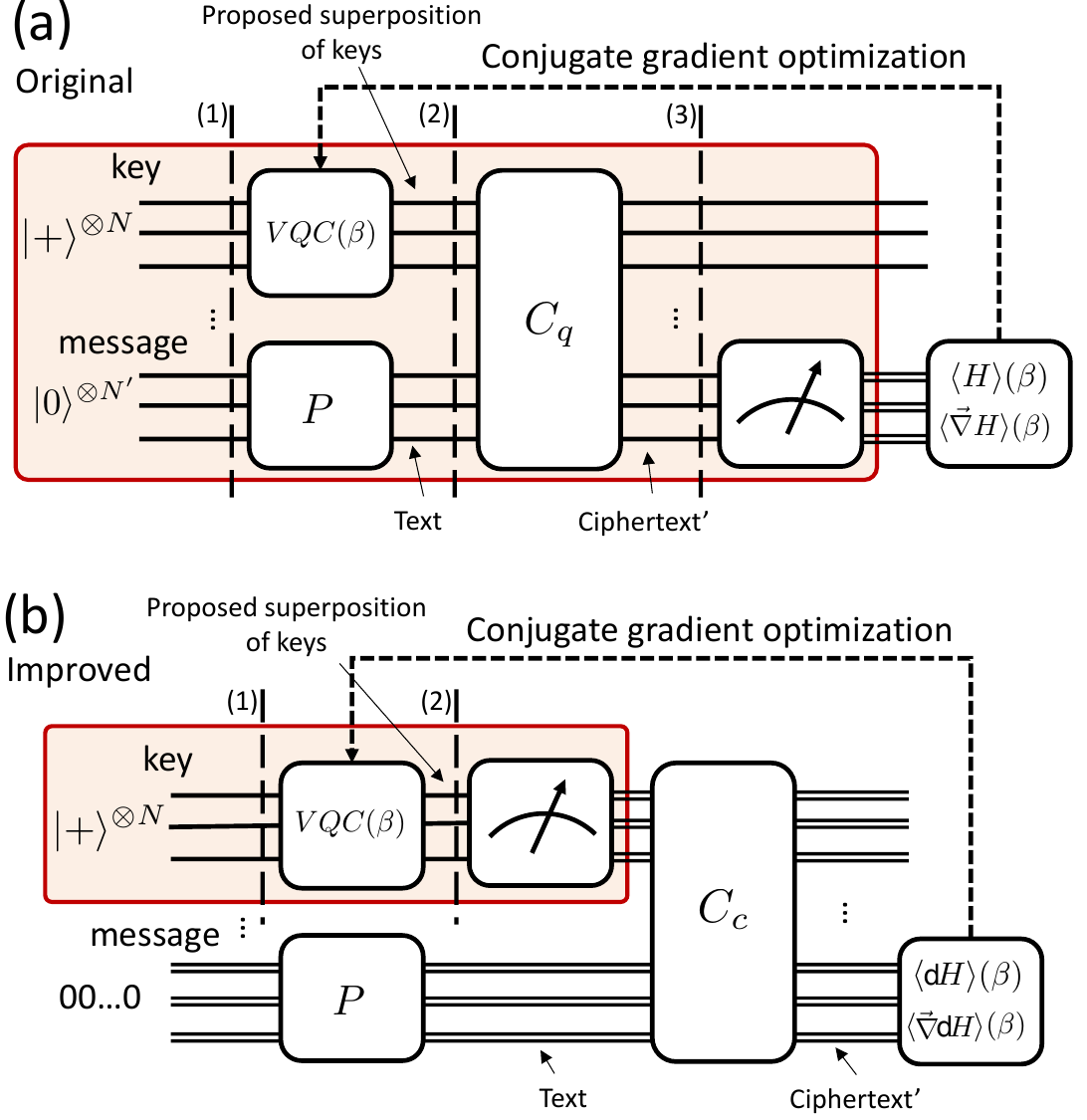}
  \caption{[Color online] Quantum circuits for (a) the original, and (b) improved VQAA algorithms, as described in the main text. Single lines correspond to qubits, and double lines correspond to classical bits. The part that runs on a quantum computer is the one inside the shaded red boxes. The improved circuit in (b) uses fewer qubits, fewer quantum gates, a big part of it is entirely classical, and produces exactly the same outcome as the original circuit in (a), which is way more expensive in terms of resources.}
  \label{Fig1}
\end{figure}

\subsection{Variational Quantum Attack Algorithm (VQAA) with Fewer Qubits}

For symmetric-key cryptography, we assume that a message is provided (plaintext), together with an encryption protocol, as well as the encrypted message (ciphertext), and the goal is then to find the encryption key.

The methodology benefits from quantum superposition and entanglement, allowing the efficient exploration of the key space while constraining the optimization landscape. These advantages are further enhanced by employing advanced variational quantum circuits (VQCs) that balance circuit expressiveness with classical simulability. As highlighted in Subsection \ref{sec_vqca}, these design considerations ensure that the algorithm remains computationally efficient and practically scalable for symmetric cryptographic problems.

\subsubsection{Original VQAA}
The Variational Quantum Attack Algorithm (VQAA) \cite{VQAA} was proposed as a quantum algorithm for Noisy Intermediate-Scale Quantum (NISQ) devices aiming at solving this problem. This is briefly schematized in the quantum circuit of Fig.(\ref{Fig1}.a). We start with a quantum computer with two registers (sections), one for the ``key'' space, with $N$ qubits, and one for the ``message'' space, with $N'$ qubits. The qubits in the message space are initialized in a reference state $\ket{0}^{\otimes N'}$, and after application of an (unitary) operation they codify some plaintext $P$. This $P$ is the message we want to encrypt, a quantum state that represents a classical string of bits. Simultaneously, the key space is initialized in a reference state $\ket{+}^{\otimes N}$, and afterwards a Variational Quantum Circuit (VQC) is applied. This quantum circuit depends on a set of variational parameters $\beta$, similar to e.g. the Variational Quantum Eigensolver (VQE) algorithm for optimizing cost functions \cite{VQE}. The VQC produces a quantum state that amounts to a (weighted) superposition of possible keys to encode $P$. Subsequently, encryption of the message is implemented by the (unitary) operator $C_q$, which produces an encoded ``Ciphertext'' in the second register for every key applied in the first register. After this, measurements are performed in the second register to read the proposed ``Ciphertext''.

The algorithm in \cite{VQAA} uses the expectation value of a classical Hamiltonian $H$ as the cost function. This expectation value represents the energy of the quantum system's state obtained from the ciphertext, allowing us to measure the discrepancy between the measured ciphertext and the known correct ciphertext. $H$ is constructed using a 3-regular graph with each of the ciphertext's 8 bits represented as a node, thus forming an 8-node graph. The Hamiltonian incorporates both pairwise interactions between connected nodes, using the Pauli-Z operator to reflect the classical bit values, and single-qubit contributions to account for individual bit states:
\begin{align}
    H & = w_{01}Z_0Z_1 + w_{06}Z_0Z_6 + w_{07}Z_0Z_7 + w_{13}Z_1Z_3  \nonumber \\
        & + w_{17}Z_1Z_7 + w_{24}Z_2Z_4 + w_{25}Z_2Z_5 + w_{27}Z_2Z_7  \\
        & + w_{34}Z_3Z_4 + w_{36}Z_3Z_6 + w_{45}Z_4Z_5 + w_{56}Z_5Z_6 \nonumber\\
        & +  \sum_{i=0}^{7}t_{i}Z_{i} . \nonumber
\end{align}

The weight term $w_{ij}$ is given by:
\begin{align}
    w_{ij}=\left\lbrace\begin{array}{c} 1 ~if~ V(i) \neq V(j), \\ -1 ~if~ V(i) = V(j). \end{array}\right\rbrace
\end{align}

and $t_i$ by:
\begin{align}
    t_{i}=\left\lbrace\begin{array}{c} 0.5 ~if~ V(i) = 1,  \\ -0.5 ~if~ V(i) = 0. \end{array}\right\rbrace,
\end{align}

where the value of the $i$-th node is represented by $V(i)$, which corresponds to the classical value (0 or 1) of the $i$-th bit. For messages exceeding 8 bits, the general approach would involve searching for the node-graph configuration that maximizes the energy difference between ground and first excited states (facilitates a more efficient convergence towards the global minimum). In our case, we employed a different cost function that scales more effectively with increasing protocol complexity. H acts on the ciphertext, which is derived from both the message and the key and the expectation value of H is then employed to optimize the variational parameters of the quantum circuit.

In short, the variational circuit VQC generates a linear combination of all possible keys. This superposition of keys serves as the input to the encryption process, which is used to produce a superposition of possible ciphertexts to the input plaintext, one for each key. The variational optimization then increases the probability of the correct key.

To further understand possible improvements, let us analyze  {the} quantum state of the circuit at different steps. In step (1) in Fig.(\ref{Fig1}.a), the overall quantum state of the system is given by 
\beq
\ket{\psi^{(1)}} = \ket{+}^{\otimes N} \ket{0}^{\otimes N'},
\eeq
where N is number of qubits for the ``key'' space (in S-DES would be 10) and N’ is for the ``message'' space (in S-DES would be 8). Subsequently, after applying the VQC in the key register as well as operator $P$ in the message register, the quantum state in step (2) is given by 
\beq
\ket{\psi^{(2)}} = \left(\sum_{k = 0}^{2^N - 1} c_k(\beta) \ket{k} \right)\ket{p},
\label{two}
\eeq
where $\ket{k}$ is the computational state for key $k$, $\ket{p}$ is the quantum state representing the plaintext $p$ obtained by applying the operator $P$ to the initial state $\ket{0}$, and $c_k(\beta)$ probability amplitudes for the key states, depending on parameters $\beta$ in the variational quantum circuit. The operator $P$ encodes the classical plaintext bits into a quantum state by starting in the $\ket{0}$ state and applying an X gate to those qubits whose corresponding classical bits are '1's.

Next, in step (3), after applying the unitary encryption operator $C_q$, we obtain the state 
\beq
\ket{\psi^{(3)}} = \sum_{k = 0}^{2^N - 1} c_k(\beta) \ket{k} \ket{p'(k,p)},
\eeq
which is generally entangled. Here, $\ket{p'(k,p)}$ is is the quantum state representing the ciphertext corresponding to key $k$ and plaintext $p$. The encryption operator $C_q$ applies the corresponding gates from the protocol to transform the plaintext into ciphertext depending on the key values. If the encryption is performed classically, we denote the operator as $C_c$, which involves classical gates, whereas $C_q$ denotes the use of quantum gates.

A measurement of the message register in this quantum state thus produces ciphertext $p'(k,p)$ with probability 
\beq
{\rm Prob}(p'(k,p)) = |c_k(\beta)|^2,
\eeq
since all states $\ket{k}$ are orthonormal. 

Therefore, at each iteration in the variational loop, the VQAA algorithm is simply sampling ciphertexts $p'(k,p)$ with probability distribution ${\rm Prob}(p'(k,p)) = |c_k(\beta)|^2$. This observation is critical for the next section, where we see how this allows to simplify dramatically the overall procedure. 

\subsubsection{Improved VQAA}

The original VQAA algorithm uses a total of $N + N'$ qubits both for the text and message registers. In addition, it has the complication that the encryption mechanism needs to be executed also on the quantum computer and, therefore, must be codified as a unitary operator $C_q$. While this is always theoretically possible and one can always design some quantum circuit for $C_q$ \cite{almazrooie2016quantum, almazrooie2018quantum, li2023new}, in practice it is not a priori obvious how to do it since encryption algorithms are naturally programmed on classical computers for everyday protocols. Moreover, adding more quantum gates implies more computational errors, given the limited capabilities of NISQ processors \cite{decoders}. 

All these shortcomings disappear by noticing that, in fact, the same outcome of the algorithm can be obtained with the quantum circuit from Fig.(\ref{Fig1}.b). The difference with the original version of VQAA, i.e., the circuit in Fig.(\ref{Fig1}.a), is that the quantum measurements are not done in the message register after applying the encryption (set of quantum gates corresponding to designed protocol), but rather \emph{in the key register before applying the encryption}. Moreover, the quantum computer is now only needed to process the VQC in the key register and, as one can see in the figure, most of the calculations \emph{are entirely classical}, including writing and encoding the message. The number of qubits is then drastically reduced to $N$ and, in addition, the encryption is entirely classical and done via some encryption protocol $C_c$ that runs on a classical computer. Or, in other words, $C_c$ is just the classical function in your favorite programming language that takes the plaintext and the key, and produces the ciphertext. These functions are already pre-programmed for all standard cryptographic protocols, and no quantum computer is needed at all for this, thus, hugely enhancing accuracy and dramatically reducing the number of required qubits. 

Let us see more in detail why the outcome of both circuits in Fig.(\ref{Fig1}) is the same. For this, we analyze the overall state of the system in the improved version in Fig.(\ref{Fig1}.b), both for the qubits in the quantum register as well as for the bits in the message register. In step (1) in the diagram, the overall register is in the state 
\beq
\ket{\psi^{(1)}} = \ket{+}^{\otimes N}  ~ [0 0 \cdots 0],  
\eeq
where only the qubits in the text register are in a quantum state, and bits in the message register are in the classical string bits $[0 0 \cdots 0]$ of $N'$ bits. Next, in step 2, after applying the variational quantum circuit in the key register and codifying the plaintext in the message register, one has the state 
\beq
\ket{\psi^{(2)}} = \left(\sum_{k = 0}^{2^N - 1} c_k(\beta) \ket{k} \right)~ [ ~p ~],
\eeq
which is equivalent to Eq.(\ref{two}), but with a classical string of bits for the second register. Now, measuring the key register in this quantum state will produce a key $k$ with probability 
\beq
{\rm Prob}(k) = |c_k(\beta)|^2. 
\eeq
The next part of the circuit is entirely classical. The encryption function $C_c$ will then codify plaintext $p$ with key $k$ with the above probability. Therefore, this function will in fact produce ciphertext $p'(k,p)$ with probability 
\beq
{\rm Prob}(p'(k,p)) = |c_k(\beta)|^2.
\eeq
Here, $p'(k, p)$ is a deterministic result of the classical encryption process $p' = C_c(k, p)$, where $k$ is the sampled key and $p$ is the plaintext. The probability ${\rm Prob}(p'(k, p))$ reflects the likelihood of the algorithm sampling the ciphertext $p'$ due to the variational parameters $\beta$ encoded in the key amplitudes $c_k(\beta)$.

Therefore, at each iteration in the variational loop, the improved VQAA algorithm is sampling ciphertexts $p'(k,p)$ with probability distribution ${\rm Prob}(p'(k,p)) = |c_k(\beta)|^2$, exactly as in the orginal proposal, but with fewer qubits, fewer quantum gates, and a big part of the method being processed entirely by a classical computer. Reducing circuit depth is specially pertinent for NISQ hardware where the noise introduced by gates can be a limiting factor of the reliable performance of a quantum algorithm \cite{decoders}.

The rest of the algorithm follows the original proposal, featuring an optimization loop via the conjugate gradient method of an appropriate cost function that measures the difference between the obtained ciphertext and the expected one. The quantum system generates keys with amplitudes $c_k(\beta)$. These keys, represented as $\ket{k}$, are applied to a classical bit string, producing the ciphertext, and the cost function is evaluated. 
While conceptually similar to Quantum Approximate Optimization Algorithm (QAOA) \cite{Farhi2014}, our method incorporates heuristics at several stages, including the design of the variational ansatz and the choice of cost functions. These heuristics are informed by practical considerations rather than systematic derivations, reflecting the empirical nature of the approach and its adaptation to cryptographic challenges.

QAOA is a hybrid quantum-classical algorithm designed to solve combinatorial optimization problems. It alternates between applying parameterized quantum operators (e.g., mixing and problem-specific unitaries) and measuring the resulting quantum states to minimize a classical cost function. In our case, the quantum computer generates bit strings (i.e., $\ket{k}$ states), which are evaluated through a classical Hamiltonian to measure the cost function. By integrating these ideas, our method facilitates efficient key generation and evaluation within the variational framework, significantly enhancing the algorithm's performance. For a detailed review of QAOA, we refer the reader to \cite{Farhi2014}.

\subsection{Non-Orthogonal Encoding}

\begin{figure}
\centering
  \includegraphics[width=1\columnwidth]{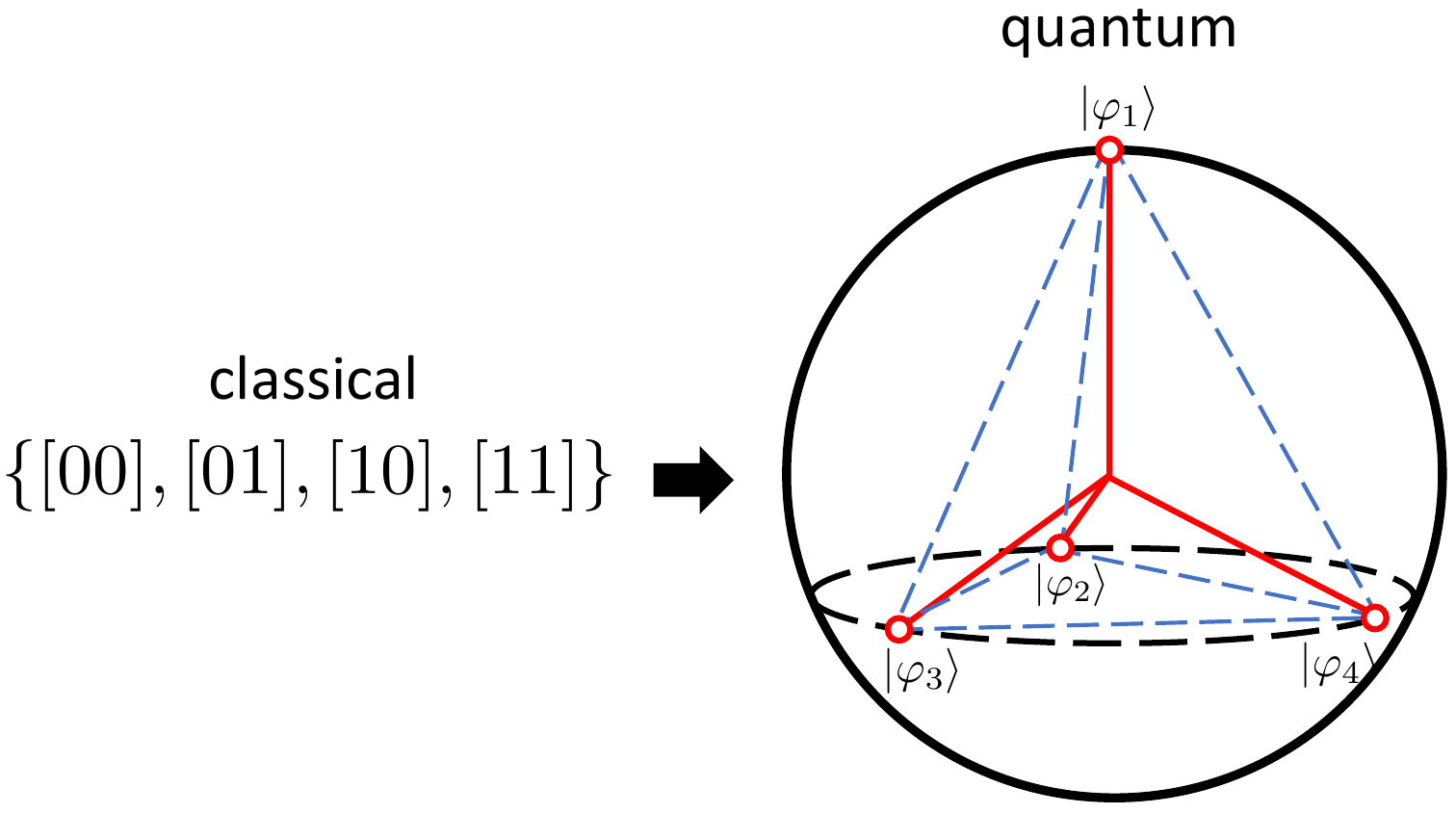}
  \caption{[Color online] The four configurations of two classical bits are mapped to four maximally close to orthogonal} states of one qubit, corresponding to the vertices of a tetrahedron inscribed in the qubit's Bloch sphere.
  \label{Fig2}
\end{figure}

Following the strategy introduced in \cite{non-orthogonal}, we also consider the possibility of using each qubit to encode more than one bit state. This method involves the use of non-orthogonal encodings, which means we select sets of states that are not perfectly distinct (orthogonal) but are designed to represent different bit configurations in a way that maximizes their distinguishability without achieving complete distinction (maximally close to orthogonal). Mathematically this can be defined as the following minimization problem:
\begin{equation}
\begin{aligned}
 \underset{A\in \mathcal{C}^{2^N\times k}}{\text{argmin}} \quad & || A^\dagger A - I_{k\times k}||_F\\
\textrm{s.t.} \quad & a_i^\dagger a_i = 1,\\
\end{aligned}
\end{equation}
where the columns of matrix $A$, i.e. $a_i$, refer to the $k$ maximally orthogonal states over $N$ qubits and $||\cdot||_F$ is the Frobenius norm. For simplicity, one of the states is usually selected to be the $\ket{0}^{\otimes N}$ state.
For instance, one could codify the four states of two bits in the following four non-orthogonal states of one qubit: 
\beq
\{[00], [01], [10], [11] \} \rightarrow \{\ket{\varphi_1}, \ket{\varphi_2}, \ket{\varphi_3}, \ket{\varphi_4} \}, 
\eeq
with the quantum states 
\beqa
\ket{\varphi_1} &=& \ket{0},  \nonumber \\ 
\ket{\varphi_2} &=& \frac{1}{\sqrt{3}} \ket{0} + \sqrt{\frac{2}{3}} e^{i \frac{2 \pi}{3}} \ket{1}, \nonumber \\
\ket{\varphi_3} &=& \frac{1}{\sqrt{3}} \ket{0} + \sqrt{\frac{2}{3}} e^{i \frac{4 \pi}{3}} \ket{1}, \nonumber \\
\ket{\varphi_4} &=& \frac{1}{\sqrt{3}} \ket{0} + \sqrt{\frac{2}{3}}  \ket{1}.  
\eeqa
The above set of four quantum states correspond to the vertices of a regular tetrahedron inscribed in the Bloch sphere of one qubit, see Fig.(\ref{Fig2}). These four states are not mutually orthogonal, but they are a set of four maximally close to orthogonal states of one qubit. In the S-DES case, this process allows to encode the 10-bit key into 5 qubits, effectively reducing the required number of qubits.

In extending the strategy introduced in \cite{non-orthogonal}, our approach enables the encoding of higher bit configurations. This reduced the qubit requirement from  n=10 to n=5 in S-DES (two classical bits into the states of one qubit), from n=16 to n=4 in S-AES (four classical bits into the states of one qubit) and from n=24 to n=6 in Blowfish (four classical bits into the states of one qubit). The resulting models run significantly faster in simulation, with only a modest dip in fidelity, demonstrating a favorable trade-off for practical applications. Specifically, for encoding higher numbers of labels, we utilize a Fibonacci sphere distribution to position states equidistantly on the sphere, ensuring maximal distinguishability among them.

For the practical implementation, this process generates equidistant points on a sphere using the Fibonacci lattice method, converts these cartesian coordinates to coefficients for quantum state representation, and appends them to a list of labels for encoding classical information into quantum states. This method results in a set of states that are not mutually orthogonal but are designed to be as distinguishable as possible within the constraints of the Bloch sphere, akin to the vertices of platonic solids for a lesser number of points.

In this implementation, we encode product states into the quantum circuit, leaving the complexity of such non-orthogonal states for the read-out stage. In order to perform such measurements, we conduct single qubit tomography (quantum state measurement). Notice that we intend here to extract local information by means of these measurements, so one could also think about implementing alternative techniques such as classical shadows \cite{Huang_2020,Huang_2022} to build local properties efficiently. This could potentially simplify the measurement protocol and extend to some hybrid solution reducing the amount of quntum resources employed in the algorithm.

To quantify the distinguishability of the encoded states during the algorithm's execution, we compute the fidelity between the expected quantum states (encoded classical configurations) and the measured states of the qubits. The fidelity calculations are based on the density matrices derived from the quantum circuit's output states. This process effectively allows us to identify the classical bit configuration that corresponds to the measured quantum state with the highest fidelity, thus decoding the encoded information.

This approach drastically reduces the number of qubits required for complex optimization problems, and enables the model to run using fewer qubits by encoding multiple classical bits of the key into a single qubit. This approach enhances the scalability of our variational algorithm, making it more feasible for implementation on NISQ devices. Moreover, it led to a significant reduction in the algorithm's simulation time, presenting a favorable trade-off even with a slight dip in performance. For further details on the implementation and potential of this encoding in variational optimization algorithms, we refer the reader to \cite{non-orthogonal}.

\subsection{Cost Function: Hamming Distance} 

The choice of an effective cost function is crucial for the success of any optimization algorithm. In the context of the algorithms discussed in this paper, the \emph{Hamming Distance} has proven to be the most effective cost function among several that were tested (which included quadratic polynomials, higher-order polynomials, $p$-norms, and more). This metric measures the difference between two binary strings $a, b$, i.e., the number of bits in which they differ. Compared to other cost functions, such as quadratic polynomials and other, we have seen that this cost function is the one that performs better in practice for this algorithm. The Hamming Distance between binary strings $a=(a_1,...,a_n)$ and $b=(b_1,...,b_n)$ is computed by summing up the bitwise XOR operations $(a_i \oplus b_i)$, i.e.
\beq
d_H(a,b) = |\{i:a_i\neq b_i\}| = \sum_{i=1}^n (a_i \oplus b_i).
\eeq

For instance, if $a = [10101]$ and $b=[11000]$, then the Hamming Distance $d_H$ is computed as 
\beq
(a \oplus b) = (10101 \oplus 11000) = 01101 \rightarrow d_H  = 3.
\eeq
 
The Hamming Distance serves as a robust measure to compare the output of the encryption block, i.e., the guessed ciphertext, with the actual true ciphertext. The algorithm iteratively updates the variational parameters based on this comparison, as we shall explain in Sec.\ref{sec2}E. A Hamming Distance of value zero indicates a perfect match between the guessed and actual ciphertexts, leading to the collapse of the key space to the correct key. In practice, we observed that this cost function happened to be the most efficient in terms of the average number of iterations needed to successfully recover the key.


\subsection{Variational Quantum Circuit Ansatz}
\label{sec_vqca}

The Variational Quantum Circuit (VQC) is the backbone of the Variational Quantum Attack Algorithm. The new ansatz introduced in this study presents a significant improvement over previous versions, offering enhanced efficiency in key space exploration. This is achieved through the inclusion of single-qubit unitary gates $U(\theta, \varphi, \lambda)$, which perform arbitrary rotations in the Bloch sphere. The matrix representation of these gates is: 
\begin{equation}
U(\theta, \varphi, \lambda) = \begin{pmatrix}
    \cos{\frac{\theta}{2}} & -e^{i\lambda}\sin{\frac{\theta}{2}} \\
    e^{i\varphi}\sin{\frac{\theta}{2}} & e^{i(\varphi + \lambda)}\cos{\frac{\theta}{2}} \\
\end{pmatrix}.
\end{equation}

\begin{figure}
\centering
\includegraphics[width=1\columnwidth]{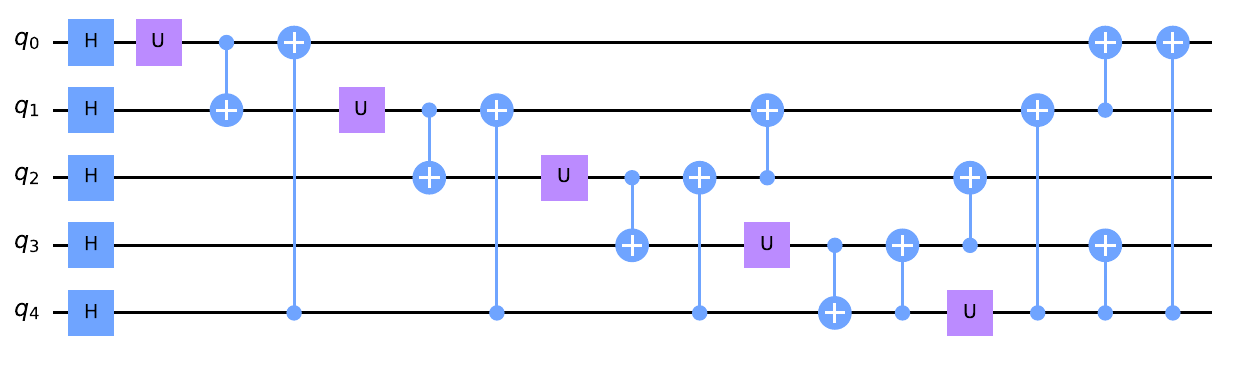}
\caption{[Color online] Variational Quantum Circuit (VQC) ansatz with a single layer, designed for the Simplified-DES encryption scheme. This circuit uses 5 qubits to encode a 10-bit key, demonstrating the encoding efficiency through non-orthogonal state representation. The configuration includes single-qubit unitary gates $U(\theta, \varphi, \lambda)$ for arbitrary rotations, complemented by a strategic arrangement of CNOT gates to enhance entanglement and optimize key space exploration.}
\label{Fig_circ}
\end{figure}

The revised algorithm for the VQAA ansatz is built as follows. The initial block of the ansatz employs Hadamard gates on every qubit to prepare each qubit in the $\ket{+}$ state, which is represented as $\ket{+}^{\otimes n} = H^{\otimes n}\ket{0}^{\otimes n}$. We also experimented with replacing the Hadamard gate layer with a $U$ gate layer with arbitrary parameters to initialize the qubits. While this approach provides additional flexibility by allowing optimization of the parameters, the performance in terms of the number of iterations required to recover the cryptographic key was similar or slightly worse compared to using the Hadamard gates.

The original paper \cite{VQAA} explored various configurations for variational parameters and entanglement gates, including CNOT, CZ, and CY gates. Building upon this, we found that using a general $U$ gate with optimized parameters for variational rotations provided modest improvements in some cases, offering a more versatile framework for the ansatz. After establishing the initial superposition, the algorithm incorporates a loop structure deploying two Controlled-NOT (CNOT) gates in each iteration: one connecting consecutive qubits, and the other linking the last qubit to the current one. This setup consistently outperformed configurations using Controlled-$Z$ gates in our experiments. The entanglement depth achieved with these CNOTs ensures extensive interaction among all the qubits. Additional CNOT gates have been integrated into the algorithm to bolster its search space by inducing further entanglement. Specifically, the control and target qubits for these gates are determined using Python's random library, with a fixed seed for reproducibility. In each outer loop corresponding to the number of layers, four CNOT gates are added. This random selection of qubits for the additional CNOTs not only enhances the entanglement but also, in some instances, leads to improved results.

As part of an exhaustive analysis to find the optimal configuration, we determined that using three layers in the variational ansatz instead of the original two, and setting the third parameter of the $U$ gates to zero, i.e., $\lambda=0$, produced the best results in terms of the average number of iterations required to successfully recover the cryptographic key.

Constructing the ansatz for our Variational Quantum Circuit (VQC) involves carefully balancing the circuit's expressiveness with its classical simulability. By expressiveness, we refer to the circuit's ability to represent complex quantum states that are challenging to simulate classically. The design aims to achieve sufficient complexity to make classical simulation inefficient—emphasizing the importance of entanglement properties that give quantum methods their advantage—while avoiding an overly expansive solution landscape that does not effectively constrain the Hilbert space, as discussed in \cite{holmes2022connecting}. This balance is crucial; if the search space is too large, it can negate the benefits of using a quantum circuit.

Additionally, recent discussions on the ability to simulate quantum circuits classically (sometimes referred to as ``classicability'') in contexts such as those outlined in \cite{cerezo2023does} highlight the need for advanced ansatz structures, strategic initialization, and other techniques to fully leverage quantum resources in cryptanalysis. Although this work primarily explores various configurations through numerical experiments—focusing on the potential of non-orthogonal state encodings and alternative coordinate systems—we recognize the need for deeper investigation into these aspects. Moving forward, we aim to extend our research to include a comprehensive review of initialization techniques and parameterized circuits, striving to refine the toolkit and search strategies for integrating quantum computing into cryptanalysis while staying attuned to the evolving landscape of quantum algorithmic complexity and its implications for cryptography.

\subsection{Coordinate Transformations for Classical Optimizations} 

As a hybrid variational approach, the optimization of the variational parameters entirely falls into the domain of a classical optimizer. In particular, the original proposal in \cite{VQAA} explored the use of Gradient Descent (GD) \cite{GD} and Nelder-Mead (N-M) \cite{NM} for updating the variational parameters of the VQC. More in depth, they point towards a slightly  modified Gradient Descent algorithm resembling the Adam optimizer \cite{Adam} by looking at the update of parameters. 

In the optimization loop, it is important to realize that each iteration is far more computationally intensive than it may initially appear. This is because each iteration consists of an inner loop that goes through the entire array of variational parameters. In this inner loop, each individual parameter is carefully adjusted, implying a recalculation of the cost function, which in turn requires performing new quantum measurements. Therefore, the gradient for each parameter is calculated based on these new measurements. At the end of the inner loop, the variational parameters are collectively updated using these gradients. It should be noted that when we refer to the number of measurements needed in each result, we mean that one measurement involves obtaining one key. The process to obtain this key involves multiple measurements to calculate the reduced density matrix (RDM) for each individual qubit. This comes from the fact that we make use of a denser encoding of the key information in each qubit by means of the previously discussed maximally non-orthogonal encoding, implying that tomography is required to know which classical register is being measured. However, note that the tomography is done independently for each of the qubits in the circuit, which can be done in parallel. Therefore, the amount of measurements that are required is independent of the number of qubits in the system, implying that the method is scalable in terms of the number of measurements. Note also that for a single iteration, we optimize each variational parameter independently and, thus, each iteration performing the previously discussed measurements of the RDM a number of times equal to the number of variational parameters. Such number has a linear dependence on the number of bits of the key, so the number of measurements per iteration will increase linearly with the key size. This linear increase of measurements is much more manageable than the exponential incresae of key combinations required for a brute force attack and, therefore, speaks well about the potential of our method to scale. This iterative nature, involving multiple measurements per iteration, makes the optimization step a crucial bottleneck in both computational resources and time, warranting close attention in future algorithmic improvements.

Our optimization strategy makes use of two distinct methods: hyperspherical coordinates and plane rotations \cite{rotaxis}. Specifically, hyperspherical coordinates improve convergence by mitigating local minima and barren plateaus, while plane rotations provide additional flexibility to navigate complex optimization landscapes. In the hyperspherical coordinates approach, the cost function $f(\vec{x})$ and original cartesian coordinates $n$ define a point \( P \) in a \( (n + 1) \)-dimensional space. The coordinates are then transformed into \( (n+1) \)-dimensional hyperspherical coordinates \( \{ \vec{\theta}, r \} \) as follows:
\begin{equation} 
    P = \left[ x_1, x_2, \cdots, x_n, f(\vec{x}) \right] \rightarrow P = \left[ \theta_1, \theta_2, \cdots, \theta_n, r \right]. 
\end{equation}
Optimization is performed on this new set of coordinates \( \{ \vec{\theta}, r \} \) with respect to the reference cost value $f(\vec{x})$. This means that we project each point from hyperspherical coordinates into the cartesian to obtain the new reference value and come back to hyperspherical coordinates where parameters are updated by changes in the original cost. The new point defined by the change $\{\vec{\theta},r\} \rightarrow \{\vec{\theta}',r'\}$ after all the parameters have been updated:  
\begin{equation}
    P' = \left[ \theta'_1, \theta'_2, \cdots, \theta'_n, r' \right]. 
\end{equation}
When convergence is reached, the final transformation $\{\vec{\theta}',r'\} \rightarrow \{\vec{x_{\rm c}},f_c(\vec{x_{\rm c}})\}$ is performed, retrieving the point P$_{\rm c}$:
\begin{equation}
     P_{{\rm c}} \equiv \left[ x'_1, x'_2, \cdots, x'_n, f'(\vec{x}') \right], 
\end{equation}
This method has shown significant improvements in the number of iterations required for convergence, making it a valuable addition to our variational quantum attack algorithms. This transformation was used in our experiments to optimize the parameters of the VQAA ansatz for the S-DES, S-AES, and Blowfish cryptanalysis, resulting in faster convergence and improved optimization stability.

The plane rotations method also starts with a point \( P \) in \( (n + 1) \)-dimensional space defined by the cost function and Cartesian coordinates. But instead of transforming to hyperspherical coordinates, a \( (n + 1) \)-dimensional rotation of the axis is performed. This is particularly useful for escaping local minima or Barren plateaus in the cost function landscape. The rotation angle can also be variable, offering flexibility in navigating the optimization landscape. Both methods extend the variables to a higher-dimensional space, including the cost function as an extra dimension. This allows for smoother and sometimes also more efficient optimizations. Plane rotations were tested as an alternative to hyperspherical coordinates for the S-DES and S-AES algorithms, demonstrating flexibility in escaping barren plateaus, although their computational overhead made them less efficient for larger keys. For further details on the practical implementation of these enhancements in the context of quantum machine learning algorithms, we refer the reader to \cite{rotaxis}.

\section{Results for Symmetric-Key Ciphers}
\label{sec3}

While our approach is completely general and can be implemented to attack essentially any type of cipher, for the sake of concreteness in this paper we focused on analyzing numerically the case of symmetric-key ciphers. In particular, we focused on the S-DES, S-AES and Blowfish protocols. Our results are discussed in the following sections.    

In our experiments and analyses, we primarily utilized Qiskit (version 0.42.1) as our quantum computing framework, running on a Python environment (version 3.9.10). The computations and simulations were conducted on a MacBook Pro 16-inch, 2021 model, which boasts an Apple M1 Pro processor and 16 GB of memory. The operating system underpinning our work was MacOS Ventura 13.0.1. All simulations were conducted on noise-free simulators to isolate the theoretical performance of the algorithm from hardware noise, ensuring a clear evaluation of the proposed approach.

\subsection{Simplified-Data Encryption Standard (S-DES)}

S-DES, or Simplified Data Encryption Standard, is a cryptographic protocol designed for secure data transmission and encryption \cite{sdes}. It is a simplified version of the widely used Data Encryption Standard (DES) algorithm, developed by IBM in the 1970s. S-DES employs a series of permutation and substitution techniques to transform plaintext into ciphertext and vice versa. Despite its simplicity, S-DES provides a fundamental understanding of symmetric key cryptography. The protocol operates on 8-bit blocks of data and uses a key length of 10 bits, making it suitable for educational purposes and basic encryption tasks. S-DES consists of a series of steps, including key generation, initial permutation, key generation for rounds, substitution and permutation operations, and a final permutation. By following these steps, S-DES ensures the confidentiality and integrity of data.

The effectiveness of our enhanced model was  assessed through numerical tests. The primary metrics for evaluation were (i) the average number of iterations needed to recover the correct key, and (ii) and the computational time for each iteration. This last figure of merit amounts, essentially, to the number of iterations multiplied by the number of operations involved in each iteration. 

To ensure a robust evaluation, we conducted 100 runs for each model, both for the original \cite{VQAA} and our improved VQAA. In each run, a different pair (plaintext, ciphertext) was used, and encrypted using S-DES. The objective was to recover the key used for encryption. For context, a brute-force attack would require an average of $2^{9} = 512$ iterations to recover the key, serving as a baseline for our quantum approach. 

\begin{figure}
\centering
\includegraphics[width=1\columnwidth]{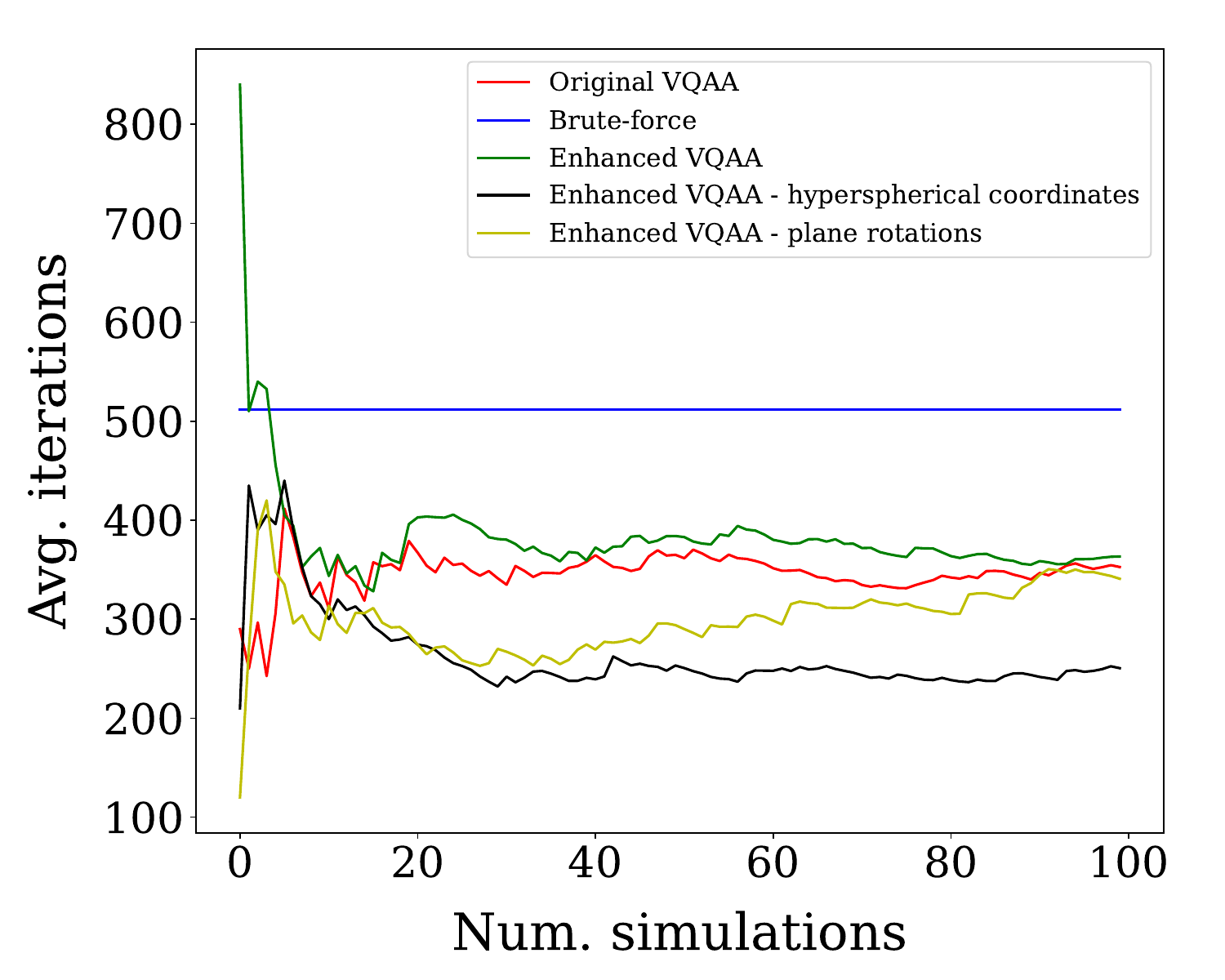}
\caption{[Color online] Comparison of the average number of iterations required to find the correct key in S-DES, using different methods: classical brute force, Original VQAA, and three different Enhanced VQAA methods (without and with coordinate transformations). The $x$-axis represents the cumulative index of simulation runs, each with different plaintexts and keys. The $y$-axis is the average number of iterations (measurements or key trials) that took to recover the correct key.}
\label{Fig3}
\end{figure}

The results of our experiment are shown in Fig.(\ref{Fig3}). As one can see, for this small key-size, the original VQAA algorithm is able to find the right key in roughly $35.27$ iterations on average, which when scaled by a factor of 10 (for the 10 measurements involved in each iteration), amounts to a total number of $352.7$ measurements on average. The enhanced VQAA shows roughly the same computational number of iterations to convergence to the correct key, which is logical since both circuits are actually different implementations of the same variational probabilistic optimization. Notice, as shown in Fig. \ref{Fig_circ}, that the enhanced version requires only 5 qubits to run, whereas the original version requires 18 qubits (10 for the key and 8 for the plaintext). The number of iterations and runtime are however decreased by adding further improvements. As shown in the figure, by adding coordinate transformations this goes down significantly, especially for the hyperspherical coordinates optimization method. This approach outperforms all others, requiring only approximately $8.3$ iterations to find the right key. When multiplied by the number of parameters involved in the VQC (30 on the whole, 3 times the circuit shown in Fig. \ref{Fig_circ}), this results in a total number of $249$ measurements on average. For comparison, the number of iterations for a classical brute-force attack is 512, which we also include in the figure. In addition, a quantum brute-force attack via Grover's quantum search algorithm would require between 18 and 25 iterations \cite{grover-attack} to recover the correct key.  

In terms of time efficiency, the hyperspherical coordinates optimization algorithm completes the $100$ simulations in roughly $134.3$ seconds, which is substantially lower than the $26,926$ seconds, nearly $7.5$ hours, required by the original VQAA model, using the hardware and software libraries mentioned earlier. To put this into perspective, this enhanced VQAA is not only faster but also more efficient in the average number of iterations needed for key recovery. Last but not least, for this small key size (10 bits) our success rate is always 100\%.

The plane rotations optimization algorithm showed on average similar performance to the Enhanced VQAA and worse than hyperspherical coordinates, but was slower due to the required matrix multiplications to transform the parameters. Thus, we focused on hyperspherical coordinates for the next cases. For S-AES and Blowfish, preliminary analyses suggested that the computational overhead of plane rotations and its similar performance to Enhanced VQAA did not justify including it in the comparative evaluation. Instead, we focused on the hyperspherical coordinates method, given its demonstrated efficiency and scalability in S-DES experiments. While this approach prioritizes practical implementation, it acknowledges that a detailed comparison for these ciphers remains a potential avenue for future work.

\subsection{Simplified Advanced Encryption Standard (S-AES)}

The Simplified Advanced Encryption Standard (S-AES) \cite{S-AES} is an educational adaptation of the widely-used Advanced Encryption Standard (AES) algorithm, more complex than S-DES. Developed to aid in understanding the complexities of AES, S-AES simplifies key components of the encryption process, making it accessible for learning and illustrative purposes. It operates on smaller data blocks, typically 16 bits, and uses shorter key lengths compared to AES, typically 16 bits. S-AES employs substitution-permutation networks, a core concept in modern symmetric key cryptography, to transform plaintext into ciphertext and vice versa. By reducing the number of rounds and employing straightforward operations, S-AES offers a practical and comprehensible entry point to study encryption algorithms. 

Unlike we did for S-DES, this time we did not test the original VQAA implementation for S-AES due to the large computational requirements involved, which made it impractical. The original VQAA algorithm would need 16 qubits for the key, plus another 16 for the plaintext, and additional qubits for the encryption process, making it hard to implement on real quantum hardware, and also more difficult to simulate with classical hardware. This is in stark contrast to our improved VQAA methods, which in this case only use at most 16 qubits, no matter the size of the plaintext.

To further reduce the qubit count, we applied a non-orthogonal encoding scheme that allowed us to use only 4 qubits (encoding 16 classical states per qubit). This approach struck a balance between efficiency and accuracy, as configurations with fewer qubits (e.g., 2 qubits encoding 256 states each) showed faster runtimes but significantly degraded performance due to poor differentiation between states. Conversely, increasing the number of qubits (e.g., 8 qubits encoding 4 states each) slightly improved accuracy but introduced substantial overhead, making it impractical from a computational perspective.

Through extensive testing, we determined that encoding 16 states per qubit (4 qubits) was the optimal threshold, achieving high accuracy while maintaining reasonable runtime. This analysis informed our choice in the Blowfish case, where we selected 16 states per qubit as the sweet spot for balancing accuracy and speed. These trade-offs demonstrate the versatility and practicality of our non-orthogonal encoding approach. The ansatz circuit for S-AES is structurally identical to the one used for S-DES, with the primary difference being that it employs 4 qubits instead of 5 and the same number of layers (3).

\begin{figure}
\centering
\includegraphics[width=1\columnwidth]{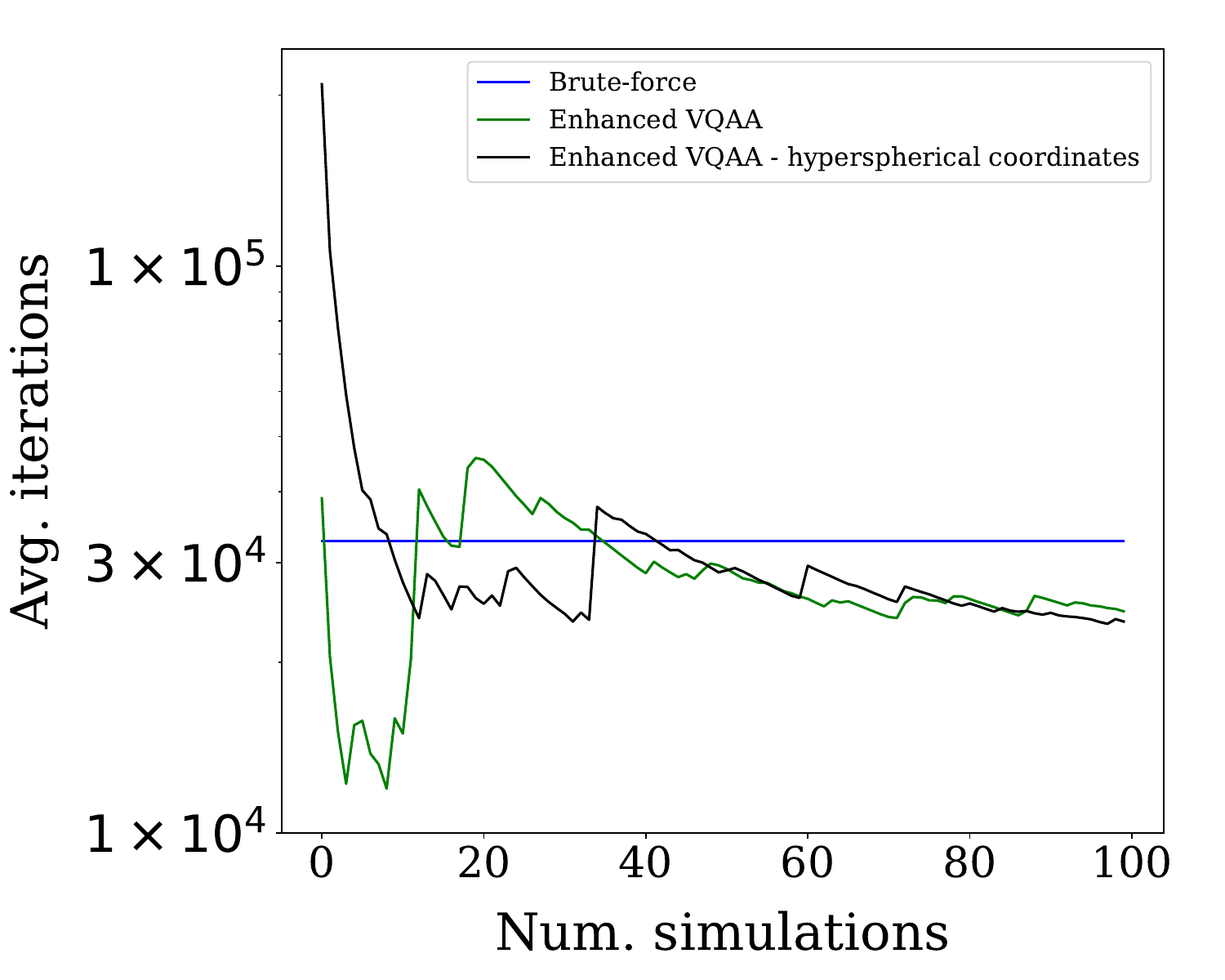}
\caption{[Color online] Comparison of the average number of iterations required to find the key using the Enhanced VQAA and classical Brute-force Attack against S-AES. The $x$-axis represents the cumulative index of simulation runs, each with different plaintexts and keys. The $y$-axis is the average number of iterations (measurements or key trials) it takes to recover the key. }
\label{fig:vqa_comp_saes}
\end{figure}

As seen in Fig.\ref{fig:vqa_comp_saes}, our enhanced VQAA algorithm shows also a compelling performance when attacking the more robust S-AES. The classical brute-force baseline for S-AES is  {$2^{15} = 32,768$} iterations. However, our hyperspherical coordinates optimization method needed approximately $982.67$ iterations, amounting to a total number of   {$23,584$} measurements on average after scaling it by a factor of 24 due to the number of variational parameters, making it therefore more efficient than brute-force. In terms of CPU time, this was achieved within an average time of about 8,241.01 seconds for all the simulations. Interestingly, similar numbers were obtained for the improved VQAA without coordinate transformations, but the convergence of this last method was slightly less smooth. 

While Grover's algorithm has been explored for S-AES in existing literature \cite{saes_grov} using 23 and 32 qubits, there are no concrete results on the number of Grover iterations required for key recovery. This makes our VQAA methods even more practical as quantum cyberattacks. Finally, our success rate is always 100\% again for this example.  

\subsection{Blowfish}

Blowfish is a symmetric-key block cipher encryption algorithm \cite{blowfish}, known for its simplicity and efficiency. It operates on fixed-size blocks of data and supports key lengths ranging from 32 bits to 448 bits, making it adaptable to various security requirements. Its key setup phase is notably fast, enabling rapid encryption and decryption processes. Despite its age, Blowfish remains widely used and respected due to its robust security features and speed. Its open design and absence of any licensing restrictions have contributed to its popularity in both commercial and open-source applications. The algorithm's resilience against various cryptanalytic attacks has solidified its reputation as a reliable choice for secure data encryption.

Quite remarkably, no effective cryptanalysis has been found to date for Blowfish, being brute force the standard attack, even though the cipher is believed to be weak against birthday attacks which, after all, are also brute force collision attacks based on the birthday paradox \cite{birthdayattack}.

Before trying a full attack on the 32-bit key, we first try a hybrid approach. For this, we fix the first 8 bits of the key to the correct ones, and reduce the search space to the remaining 24 bits. As such, this approach is not scalable, but reduces significantly the configuration space and allows us to do a systematic analysis via our simulations. As a matter of fact, this analysis is also not unrealistic: it amounts to running our improved VQAA methods in parallel in $2^8 = 256$ few-qubit quantum processors. As we shall see, we find sufficient to use $6$ qubits to run our approach, so this would amount to $256$ independent 6-qubit quantum processors. This is realistic and plausible with current technology, considering that current NISQ processors already handle thousands qubits with, e.g., neutral atoms \cite{atom}. Similar to the ansatz used for S-DES and S-AES, the ansatz for Blowfish maintains the same overall structure but is scaled to use 6 qubits instead of 5 or 4. Additionally, only a single layer is required instead of 3.

\begin{figure}
\centering
\includegraphics[width=1\columnwidth]{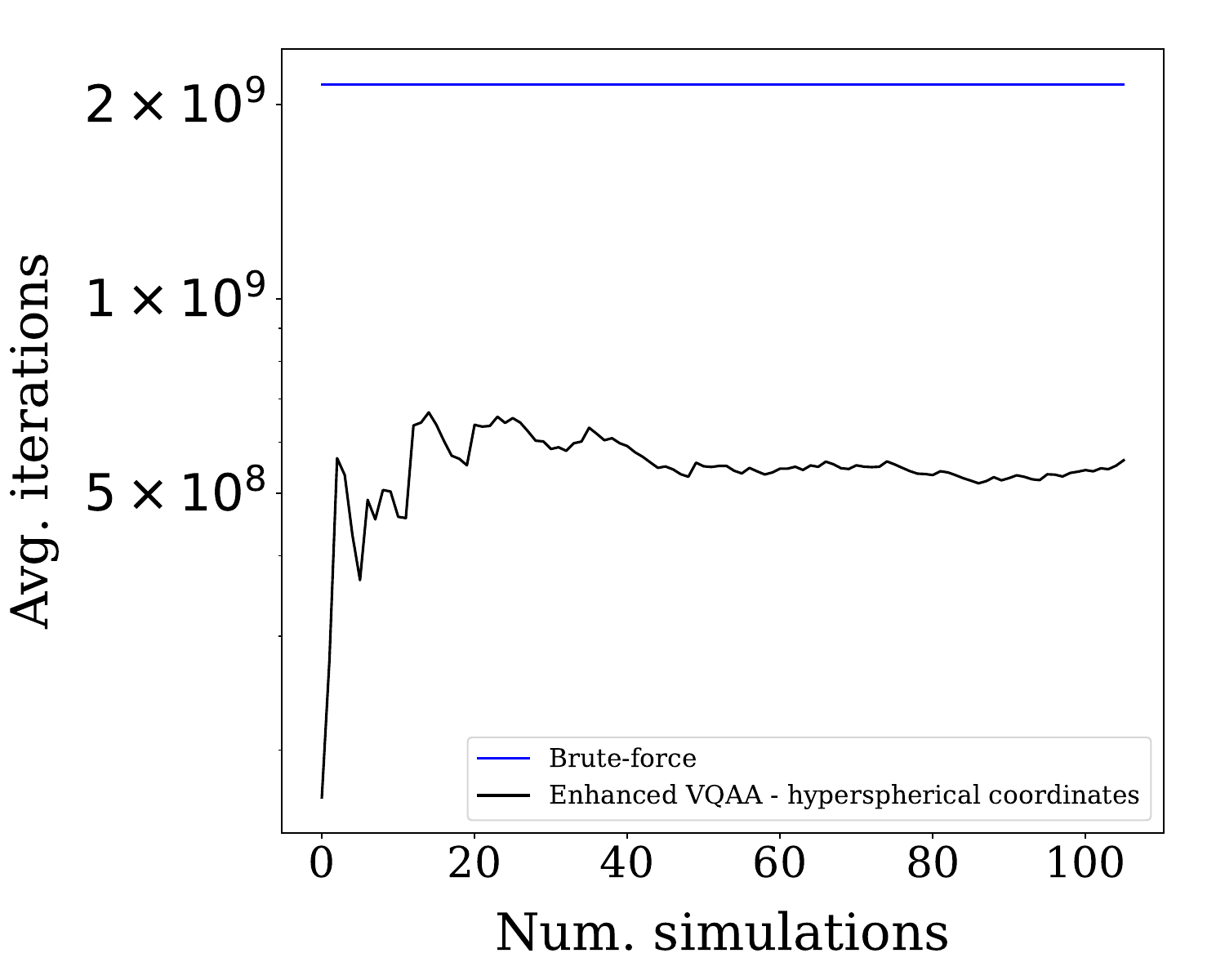}
\caption{[Color online] Comparison of the average number of iterations required to find the key using the Enhanced VQAA and Brute-force Attack against Blowfish. The x-axis represents the cumulative index of simulation runs, each with different plaintexts and keys. The y-axis is the average number of iterations (measurements or key trials) it takes to recover the key.}
\label{Fig5}
\end{figure}

For this hybrid experiment, we find an average of $365,912.7$ iterations for our improved VQAA approach (with hyperspherical coordinates) to recover the remaining 24 bits. Taking into account that we used 12 variational parameters, this translates to an average total number of $4,390,952.4$ measurements. If we would like to get rid also of the parallelisation (say, imagine that we have access to just one quantum processor), then we must multiply this by $128$, which is the average number of iterations to recover the first 8 bits, so that the total average number of measurements becomes $562,041,907.2$, see Fig.(\ref{Fig5}). Notice, importantly, that by running things sequentially and not in parallel we could reduce significantly the standard deviation of our approach by, e.g., fine-tuning the unitaries in the VQC, so that we could increase sequentially the chances of getting the correct bits, therefore reducing the overall variance of the result. Remarkably, this is significantly less than $2^{31} = 2,147,483,648$, which is the average number of iterations required for a brute-force attack to recover the entire key, as it can be seen in Fig.\ref{Fig5}. As in the previous examples, our success rate is always 100\% in this analysis. 

As a proof of principle, this hybrid cryptanalysis method for Blowfish fills a significant gap in the literature and demonstrates the capability to recover the key in far fewer iterations compared to the existing brute-force methods, which are currently considered the best-known cryptanalysis techniques for Blowfish. The gap is even larger if we allow for parallelisation of the search. 

As a further test, one of our experiments achieved a significant milestone. In a separate set of simulations of our improved VQAA method looking for the whole 32-bit key, our algorithm succeeded in finding the correct 32 bits in at least one instance. To put this in perspective, while the brute force approach demands an average of $2^{31} = 2,147,483,648$ iterations, our quantum framework reduced this to a mere number of $89,460,336$ measurements, amounting to a formidable reduction by a factor of 24. This was implemented using $1,863,757$ iterations of an 8-qubit quantum circuit, non-orthogonal encoding, and 3 layers of unitary gates with 48 variational parameters. We take this result as an even stronger proof of principle of the validity of our approach. This successful key-finding was computationally very intensive and does not give us any significant statistics. But we believe that all this can be accelerated with further improvements on the algorithm, which we are currently investigating, and with the potential to unlock increasingly complex protocols. Some of our ideas in this respect are discussed in the next sections. 

Before moving beyond symmetric-key cryptanalysis, Table~\ref{tab:comparison} provides a summary of our results, detailing the number of qubits, required quantum gates, iterations, and execution times for each method. This comparative overview highlights the efficiency of our enhanced VQAA approaches in reducing the number of iterations compared to brute-force attacks.

\begin{table*}[htbp]
    \centering
    \caption{Comparison of Quantum Algorithms with Classical Brute Force}
    \label{tab:comparison}
    \begin{tabular}{|c|c|c|c|c|c|c|c|}
        \hline
        \textbf{Algorithm} & \textbf{Qubits} & \textbf{U Gates} & \makecell{\textbf{CNOT} \\ Gates} & \makecell{\textbf{VQAA} num.\\ iterations} & \makecell{\textbf{Brute-force}\\ num. iterations} & \makecell{\textbf{VQAA} \\Runtime (s)} & \makecell{\textbf{Brute-force}\\ Runtime (s)} \\ \hline
        S-DES     & 5      & 15             & 42         & 249             & 512                & 134.3             & 1.25               \\ \hline
        S-AES     & 4      & 12             & 36         & 23,584          & 32,768             & 8,241.01          & 13.48              \\ \hline
        Blowfish  & 6      & 6              & 16         & 4,390,952.4     & 8,388,608          & 86,051            & 15,558             \\ \hline
    \end{tabular}
\end{table*}

\section{Beyond Symmetric-Key Protocols}
\label{sec4}

So far we have discussed how to apply our variational quantum method to design attacks to symmetric-key ciphers. The idea, however, is generic, and can be well implemented beyond symmetric keys. In the following we describe briefly two examples: asymmetric key protocols, such as RSA, and Hash functions, such as those used in cryptocurrencies. In both cases one can apply improved VQAA attacks, as we shall see.  

\subsection{Asymmetric-Key Cryptographic Protocols}

Asymmetric key cryptographic protocols, also known as public-key cryptography, revolutionized the field of secure communication by introducing a groundbreaking concept: the use of two distinct but mathematically related keys – a public key and a private key. Unlike symmetric key algorithms, where the same key is used for both encryption and decryption, asymmetric key protocols leverage a pair of keys to enhance security and enable secure data exchange over insecure networks like the internet.

In this innovative approach, the public key is freely shared and can be accessed by anyone, while the private key is kept confidential and known only to the intended recipient. The strength of asymmetric key cryptography lies in the computational complexity of deriving the private key from the public key, making it practically infeasible for unauthorized parties to decipher encrypted information.

This technology underpins various essential security mechanisms, including digital signatures, secure email communication, and secure web browsing through protocols like Hypertext Transfer Protocol Secure (HTTPS). Asymmetric key protocols facilitate secure key exchange between parties who have never met before, ensuring the confidentiality and integrity of data transmissions in an increasingly interconnected digital world. 

\begin{figure}
\centering
\includegraphics[width=1\columnwidth]{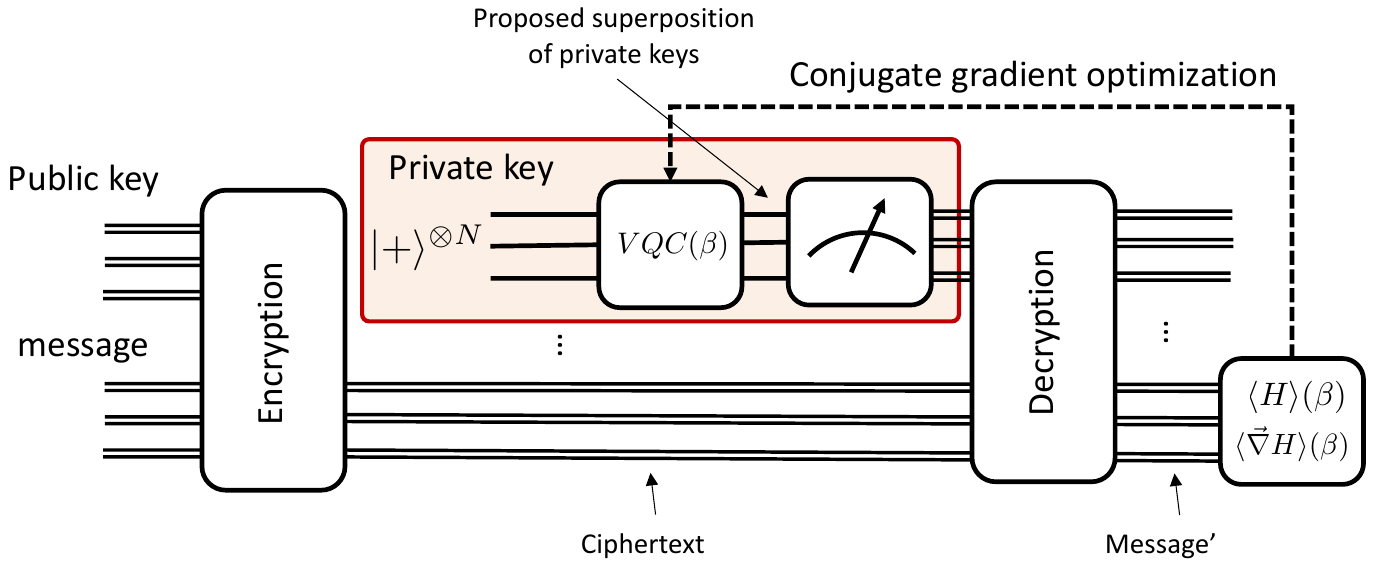}
\caption{[Color online] Quantum circuit for improved VQAA algorithm to hack asymmetric-key cryptography, as described in the main text. Single lines correspond to qubits, and double lines correspond to classical bits. The part that runs on a quantum computer is the one inside the shaded red boxes. The circuit uses few qubits, few quantum gates, and a big part of it is entirely classical.}
\label{Fig6}
\end{figure}

An improved variational quantum attack to asymmetric key cryptographic protocols is described in Fig.(\ref{Fig6}). The attack is similar to the one we implemented for symmetric keys. In this case, given a message, a ciphertext, and the public key, the goal is to obtain the private key, which is the one used for decryption. As shown in the figure, the quantum computer runs only on a register of qubits for the private key, where the VQC is applied. Measurements are implemented at the end of the circuit, and the rest of the protocol is entirely classical, including encryption, decryption, and the iterative optimization loop. 

\subsection{Hash Functions}

Hash functions are fundamental cryptographic tools that play a pivotal role in ensuring data integrity, security, and authenticity (proof of origin) in various digital applications, including cryptography and cryptocurrencies. A hash function is a mathematical algorithm that takes an input (or message) and produces a fixed-size string of characters, which is typically a unique representation of the input data.

In cryptography, hash functions are employed to verify the integrity of data by generating a unique hash value for a given input. Even a minor change in the input data results in a substantially different hash value (avalanche effect), making it computationally infeasible for attackers to tamper with the original data without detection. This property is vital in password storage, digital signatures, and data verification processes, ensuring the authenticity of information in secure communication.

\begin{figure}
\centering
\includegraphics[width=0.75\columnwidth]{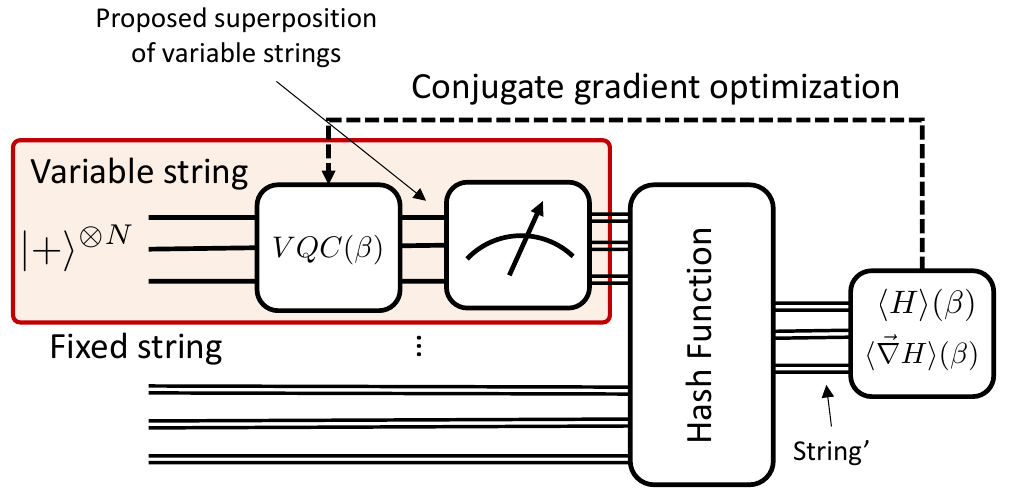}
\caption{[Color online] Quantum circuit for improved VQAA algorithm to hack hash functions, as described in the main text. Single lines correspond to qubits, and double lines correspond to classical bits. The part that runs on a quantum computer is the one inside the shaded red boxes. The circuit uses few qubits, few quantum gates, and a big part of it is entirely classical.}
\label{Fig7}
\end{figure}

In the realm of cryptocurrencies, such as Bitcoin and Ethereum, hash functions are integral to the blockchain technology. Each block in a blockchain contains a hash of the previous block, creating a chain of blocks linked together through these hash values. This linkage ensures the immutability of the blockchain, making it extremely difficult to alter past transaction records without changing all subsequent blocks, a feat that demands enormous computational power.

The hack of hash functions is related to the so-called \emph{collision problem} in mathematics. The collision problem refers to the challenge of finding two different inputs that produce the same hash output. In other words, it involves discovering two distinct pieces of data that, when processed through the hash function, result in identical hash values. Collision resistance is a crucial property for hash functions, ensuring that it is computationally infeasible to find such input pairs intentionally. If collisions are found for a hash, the hash is no longer considered as secure.  

In Fig.(\ref{Fig7}) we show an improved variational quantum attack to hash functions, targeting at finding collisions to a given hash. For this, let us assume that we have a string of bits representing a message, say, a document, with a given hash. Our goal is to find an alteration of part of the document that produces the same hash as the original one, therefore finding a collision and thus making the hash invalid. For this, we promote the bits representing the part to be altered to qubits. As shown in the figure, the variational quantum circuit and the measurements are implemented on that part of the input, which is then processed by the hash function together with the fixed bits of the document. The produced hash is then used to measure a cost function with respect to the original hash, and then proceed with the iterative optimization loop as usual. 

\section{Further improvements}
\label{sec5}

The developments in this paper show a great promise of variational quantum attacks for cryptographic protocols. What is even more exciting, is that the algorithms presented and discussed here can still be improved in a number of ways, therefore bringing them closer to hacking large cryptographic ciphers. Some further enhancements are the following.  

\begin{enumerate}

\item {\bf Symmetry of Symmetric-Key Protocols}: the attacks that we implemented here for symmetric-key protocols such as S-DES, S-AES and Blowfish, did not use at all the symmetry between encryption and decryption. This symmetry introduces correlations that could be used to the advantage of a hacking protocol, as shown in the example of Fig.(\ref{Fig8}). We expect such attacks to be more accurate and efficient for symmetric-key cryptography. 

\begin{figure}
\centering
\includegraphics[width=1\columnwidth]{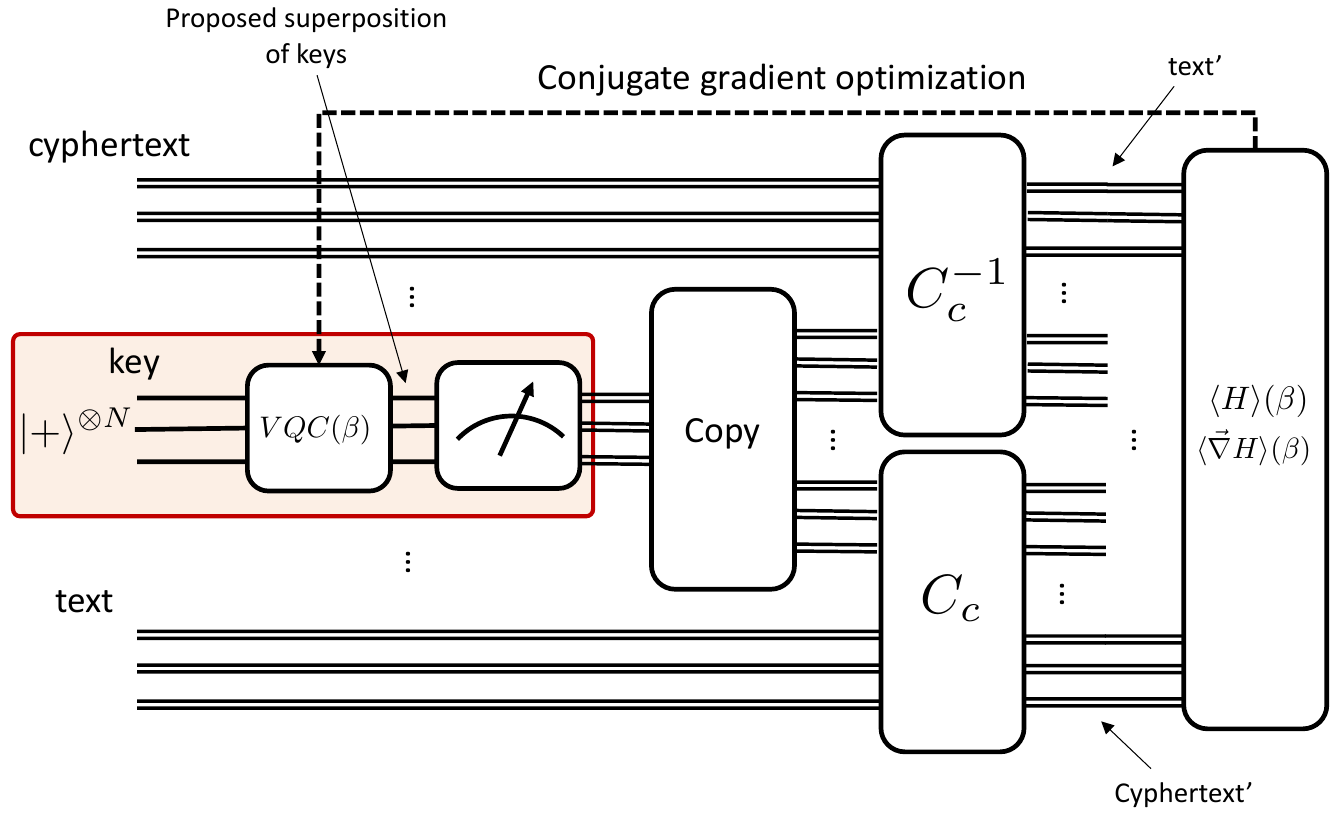}
\caption{[Color online] Quantum circuit for improved VQAA algorithm to hack symmetric-key cryptographic protocols, using the symmetry between encryption and decryption to our advantage. Single lines correspond to qubits, and double lines correspond to classical bits. The part that runs on a quantum computer is the one inside the shaded red boxes. The circuit uses few qubits, few quantum gates, and a big part of it is entirely classical. The main difference with the circuit in Fig.(\ref{Fig1}b) is that, this time, the output of the quantum measurement, which is a string of classical bits, is copied and goes as the input of (i) encryption with $C_c$, and (ii) decryption with $C_c^{-1}$. The cost function measures the offset between the outputs for both encryption and decryption and the correct text and ciphertext. In this case, the cost function can also be adapted to include correlations between text and ciphertext.}
\label{Fig8}
\end{figure}

\item \textbf{Batch Processing and Tree-Structured Optimization}: for the classical simulations, batch processing serves as a divide-and-conquer method, allowing parallel execution across multiple CPUs. Tree-structured optimization aids in efficiently navigating the key search space by starting the optimization process from favorable local minima.
    
\item \textbf{Bit-level Correlations}: it would also be possible to design cost functions that are sensitive to different types of correlations between the bits. This also motivates the exploration of more complex structures within the quantum circuit to enhance the update of variational parameters.
    
\item \textbf{Local Energy Minima Encoding}: why looking for the global minima of the cost function? In practice, we could also encode the cryptographic key in a local minima, thus circumventing challenges such as other local minima and barren plateaus. The key could then be obtained via sampling, and the procedure could work depending on the degeneracy of the local minima for the state of the key (variability in solutions).  

\item \textbf{Dual Optimization}: for symmetric protocols, by considering the percentage of relevance between cipher and plaintext, one would be able to  focus more on either the encryption or the decryption process, providing additional clues towards optimization.

\item \textbf{Embedded Variational Parameters}: this involves incorporating variational parameters directly within the encryption block, though constrained by higher qubit requirements and computational costs.
    
\item \textbf{Shotgun Sequencing}: This strategy, inspired by genetics, involves generating and comparing multiple incorrect ciphertexts and keys with the correct versions to identify any underlying patterns or actionable insights. The correct key could then be inferred by a variety of methods, including deep learning. 

\item \textbf{Tensor Networks:} the approaches developed here rely on using quantum computers to produce variational states. But why using quantum computers at all? Instead of the variational quantum circuit, one could use some variational tensor network ansatz \cite{TN}, and develop an entirely-classical variational tensor network attack. The more complex the tensor network, the more complex the attack.

\end{enumerate}

\section{Conclusions}
\label{sec6}

In this paper we have developed improved variational quantum attacks to cryptographic protocols. We applied these attacks in practice to symmetric-key ciphers, such as S-DES, S-AES and Blowfish. Classical simulations of our attacks, conducted using noise-free Qiskit simulators, allowed to find the 32-bit secret key of Blowfish. Our results demonstrate the ability of variational quantum attacks to perform slightly better than brute-force attacks on certain cryptographic protocols. Specifically, our improved VQAA methods achieve reductions in the number of required iterations and measurements, as shown in the comparative analysis discussed earlier.

While our experiments focused on simplified encryption algorithms with reduced key spaces, such as a 24-bit search space for Blowfish after assuming knowledge of the first 8 bits, scaling these attacks to larger key sizes such as AES-128 would require approximately 30 qubits. At this scale, quantum simulators become impractical due to exponential runtime and memory requirements. Using real quantum hardware for such attacks presents additional challenges, including noise, decoherence, and the limited fidelity of current devices. Furthermore, although VQAA offers performance improvements over brute-force attacks—potentially by factors of 2, 4, or even 8—cracking a 128-bit key would still involve an astronomical number of trials, requiring resources that are far beyond current or foreseeable hardware capabilities.

Nevertheless, our work highlights the foundational potential of variational quantum attacks to challenge traditional cryptographic standards. It demonstrates the scalability of VQAA with the number of measurements per optimization iteration scaling linearly with the key size. While cracking AES-128 or other modern cryptosystems is currently infeasible, these results serve as a stepping stone for future research into quantum and quantum-inspired cryptanalysis. As quantum processors improve in error rates, qubit counts, and coherence times, the applicability of these techniques could extend to more complex cryptosystems, emphasizing the need for robust quantum-resistant cybersecurity standards \cite{errormitigation, BernsteinPQC,PQCreview}.

{\bf Acknowldgements.-} We acknowledge Donostia International Physics Center (DIPC), Ikerbasque, Basque Government, Diputaci\'on de Gipuzkoa and European Innovation Council (EIC) and Tecnun for constant support, as well as insightful discussions with the teams from Multiverse Computing, DIPC and Tecnun on the algorithms and technical implementations. This work was supported by the Spanish Ministry of Science and Innovation through the project ``Few-qubit quantum hardware, algorithms and codes, on photonic and solid-state systems''  (PLEC2021-008251) and by the Diputaci\'on Foral de Gipuzkoa through the ``Quantum error mitigation for near-term quantum computers '' project (IS172551023).

\bibliography{biblio.bib}

\end{document}